\documentclass[journal=jpcafh,manuscript=article]{achemso}
\usepackage[version=3]{mhchem} % Formula subscripts using \ce{}
\usepackage[T1]{fontenc}       % Use modern font encodings

\usepackage{textgreek} % for \textbeta
\usepackage{subcaption}
\usepackage{textcomp} % for \textminus

\usepackage{todonotes}
\usepackage{float}
\author{Pierre L. Bhoorasingh} % bhoorasingh.p@husky.neu.edu
\affiliation[Northeastern University]
{Department of Chemical Engineering, Northeastern University, Boston MA 02115, USA}
\alsoaffiliation{Current address: PillPack Inc., Somerville MA}
\author{Belinda L. Slakman}% slakman.b@husky.neu.edu
\affiliation[Northeastern University]
{Department of Chemical Engineering, Northeastern University, Boston MA 02115, USA}
\author{Fariba Seyedzadeh~Khanshan}% seyedzadehkhanshan.f@husky.neu.edu
\affiliation[Northeastern University]
{Department of Chemical Engineering, Northeastern University, Boston MA 02115, USA}
\alsoaffiliation{Current address: Novartis Institutes for BioMedical Research, Cambridge MA}
\author{Jason~Y.~Cain}% cain.ja@husky.neu.edu
\affiliation[Northeastern University]
{Department of Chemical Engineering, Northeastern University, Boston MA 02115, USA}
\alsoaffiliation{Current address: Department of Chemical and Biological Engineering, Northwestern University, Evanston IL}
\author{Richard H. West}% r.west@northeastern.edu
\email{r.west@northeastern.edu}
\affiliation[Northeastern University]
{Department of Chemical Engineering, Northeastern University, Boston MA 02115, USA}

\title[Automated TST calculations]
  {Automated transition state theory calculations for high-throughput kinetics}


\begin{document}

%\begin{tocentry}

% Some journals require a graphical entry for the Table of Contents.
% This should be laid out ``print ready'' so that the sizing of the

% Inside the \texttt{tocentry} environment, the font used is Helvetica
% 8\,pt, as required by \emph{Journal of the American Chemical
% The surrounding frame is 9\,cm by 3.5\,cm, which is the maximum
% permitted for  \emph{Journal of the American Chemical Society}
% graphical table of content entries. The box will not resize if the
% content is too big: instead it will overflow the edge of the box.

% This box and the associated title will always be printed on a
% separate page at the end of the document.
%\vfill
%\includegraphics[width=5cm]{TOC-pes.pdf}
%\vfill
%\end{tocentry}

\begin{abstract}
A scarcity of known chemical kinetic parameters leads to the use of many reaction rate estimates, which are not always sufficiently accurate, in the construction of detailed kinetic models.
To reduce the reliance on these estimates and improve the accuracy of predictive kinetic models, we have developed a high-throughput, fully automated, reaction rate calculation method, AutoTST. 
The algorithm integrates automated saddle-point geometry search methods and a canonical transition state theory kinetics calculator.
The automatically calculated reaction rates compare favorably to existing  estimated rates.
Comparison against high level theoretical calculations show the new automated method performs better than rate estimates when the estimate is made by a poor analogy.
The method will improve by accounting for internal rotor contributions and by improving methods to determine molecular symmetry.

\end{abstract}

\section*{Introduction}
Detailed chemical kinetic modeling of complex systems has been aided by software that automatically generates reaction mechanisms\cite{VandeVijver:2015ba, Blurock:2013et}.
One example of such software, Reaction Mechanism Generator (RMG)\cite{Gao:2016dk}, has been applied to systems such as the pyrolysis and combustion of isobutanol\cite{Merchant:2013kz}, the fast pyrolysis of bio-oil\cite{SeyedzadehKhanshan:2016ga}, and the auto-oxidation of a biofuel surrogate\cite{BenAmara:2013kl}.
Mechanism generators can help ensure that important reaction pathways are not missed, but as a result they require thousands or even millions of thermodynamic and kinetic parameters to complete the model construction. 
These parameters are preferentially sourced from experimental measurements or accurate theoretical calculations, but most of the required parameters are unknown, leading to frequent use of less accurate estimates \cite{Broadbelt:2005ez}.

Parameter estimation methods provide thermodynamic and kinetic values in a computationally efficient manner\cite{Yu:2004cd}.
Estimation methods are typically based on Benson's group additivity method for thermochemistry\cite{Benson:1969gq}, in which group values are first determined from molecules with known thermodynamics, then used to estimate the thermodynamics of other molecules.
Benson's group contributions have been used to make adequate thermochemistry predictions for a variety of systems, including hydrocarbons\cite{Sumathi:2002jv,Sebbar:2003gk} and silicon hydrides\cite{Wong:2004gl,Adamczyk:2011ex}.
Despite these successes, group contribution methods have been difficult to extend to some cases, such as predicting thermodynamics for polycyclic species, where the ring strain causes the molecule to be poorly described by the sum of its parts. 
The RMG software addresses this deficiency in the group additive approach by performing semi-empirical or quantum mechanical calculations of thermodynamic parameters for polycyclic species\cite{Magoon:2012hg}.

For  estimating reaction kinetics, the Evans-Polanyi relationship is a simple approach in which the change in enthalpy is used to predict the kinetics of the specific reaction\cite{Evans:1936ti}.
%However, it is not always appropriate to apply the Evans-Polanyi relationship, such as in the hydrogen abstraction by methyl radicals from polynuclear aromatics\cite{KarenHemelsoet:2004hk}.
Alternative approaches extend group contribution methods to predict kinetic parameters\cite{Sumathi:2002ii,Saeys:2004ko,Adamczyk:2009es,West:2011ur, Vandeputte:2012hu}. % removed: Saeys:2003ku, Saeys:2006gw, Adamczyk:2010ds
Group estimation methods can be automated efficiently, making them useful for mechanism generators where specific reaction rates are often unavailable\cite{Yu:2004cd}.
Group-based predictions can be further improved using a hierarchy of reaction rate rules for increasingly specific reacting functional groups\cite{Curran:1998bx,Curran:2002kl,Carstensen:2009hl,Villano:2013kr}. 
Unfortunately, appropriate rate rules are rarely available when studying new systems. 
In these situations more general (less specific) rules are used, and the accuracy of the estimates suffers; sometimes the estimates resulting from these generic or inappropriate rules can be wrong by several orders of magnitude.

Continuing advances in computing power have increased the feasibility of replacing these estimates with more accurate theoretical calculations using transition state theory (TST) and quantum chemistry methods.
However, such calculations currently require a great deal of human input, mostly in locating the transition state, preventing their use in a high-throughput manner.
Automating transition state calculations in order to calculate unknown kinetic parameters has been identified by the US Department Of Energy as a basic research need for clean and efficient combustion of 21st century transportation fuels\cite{McIlroy:2006js}, and the Combustion Energy Frontier Research Center as an ``important grand challenge''\cite{Law:2010ut}.

Many parts of this challenge have been at least partially automated.
TST calculations require the three-dimensional structures of the reactants, products, and transition state involved in each reaction.
Reactant and product structures can already be found using the automated software integrated in RMG to calculate species thermochemistry\cite{Magoon:2012hg},  using a combination of distance geometry, force fields, and semi-empirical quantum chemical calculations, to propose, optimize, and compare many conformer geometries for each compound.
The artificial force induced reaction (AFIR) method\cite{Maeda:2009kg,Maeda:2014jq}, the KinBot software\cite{Zador:un}, and other methods\cite{Zimmerman:2013bb,Zimmerman:2014ko,Zimmerman:2015dp,Rooks:2014kz,Bhoorasingh:2015dza} can use computational chemistry to automatically locate the necessary transition state geometries.
Kinetic programs such as CanTherm\cite{CanTherm:2016}, VariFlex\cite{klippensteinvariflex},  MultiWell\cite{baker2012multiwell}, and POLYRATE\cite{1992CoPhC..71..235L} have been developed to calculate reaction kinetics if provided the quantum chemistry outputs.
Integrating geometry search software with kinetic calculators is a promising route to enable high-throughput kinetics calculations.

\begin{figure}[H]
\centering	
\includegraphics[width=11cm]{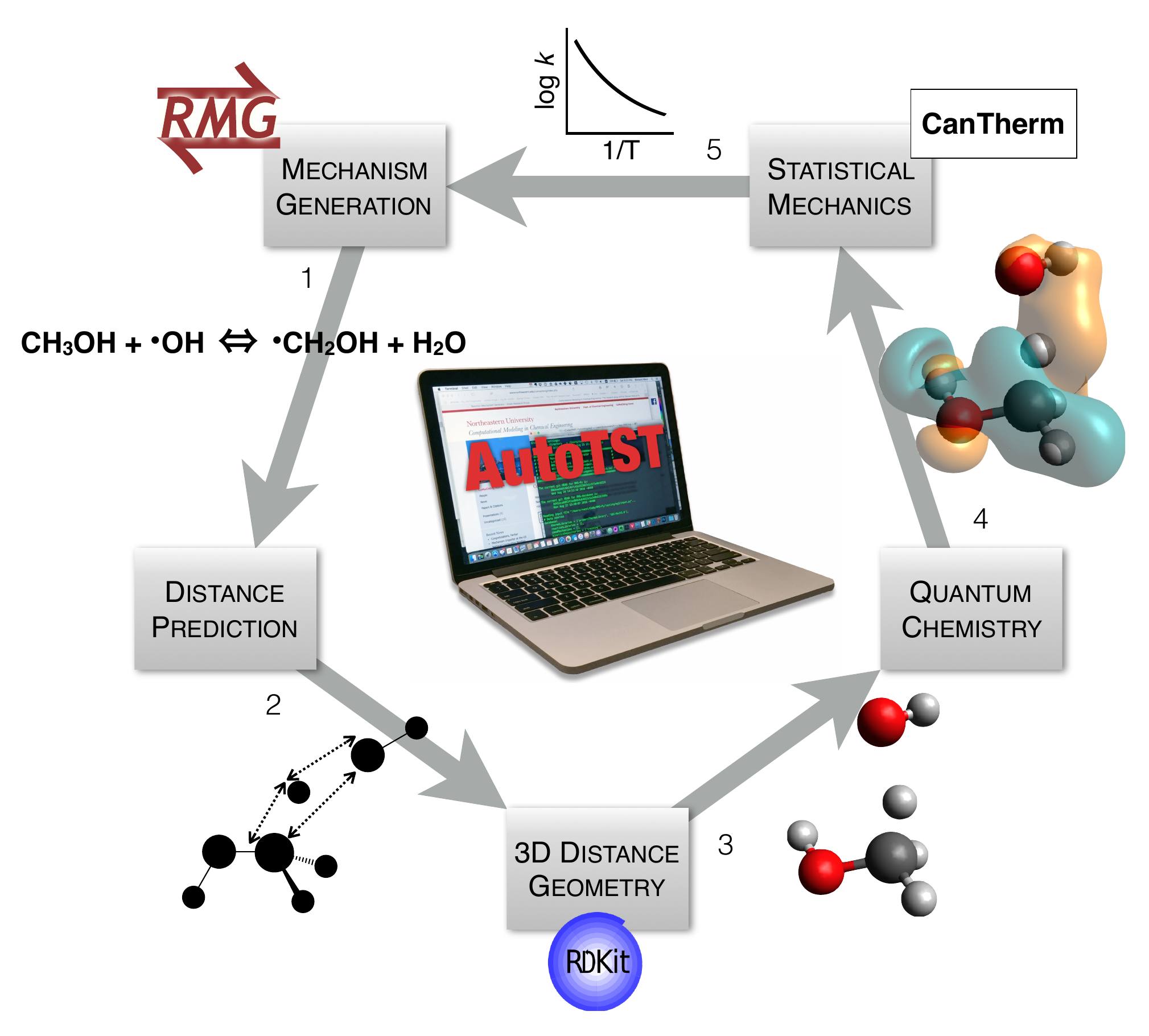}
\caption{\label{fig:overview} Overview of the AutoTST algorithm.}
\end{figure}

The present article describes automated algorithms to locate reactants, products, and transition states based on distance geometry\cite{Magoon:2012hg,Bhoorasingh:2015dza} and their integration with the CanTherm\cite{CanTherm:2016} code, and Reaction Mechanism Generator (RMG) software\cite{Gao:2016dk} to calculate reaction rate expressions.
The integrated algorithm (Fig.~\ref{fig:overview}) is referred to as the Automated Transition State Theory (AutoTST) calculator. 
Full details of the algorithms are provided in the Methods section below. %%%SECTION NUMBER?%%%
In this work we demonstrate the method on three families of reactions, chosen to cover the three main classes of molecularity: 
intramolecular H transfer or migration which is unimolecular isomerization (1 reactant, 1 product), radical addition to a multiple bond which is addition (2 reactants, 1 product), and H abstraction which is bimolecular (2 reactants, 2 products). 
The reverse of radical addition to a multiple bond is \textbeta-scission, which is unimolecular decomposition (1 reactant, 2 products).
Although there are many more reaction families, these three demonstrate the generality of the approach and together comprise about half of the reactions in typical combustion models (eg. 48\% for the butanol model described below). 
We posit ring opening (and closing) will be algorithmically quite like intramolecular H transfer in terms of determining the 3D geometry of the transition state, although the electronic structure calculations will require additional considerations to find accurate barrier heights.

Many important reactions in high temperature gas phase systems such as combustion are pressure-dependent, either chemically activated or in the fall-off regime. 
%The Reaction Mechanism Generator software\cite{Gao:2016dk} 
The RMG software contains procedures to automatically perform master equation calculations for these unimolecular reaction networks to calculate phenomenological $k(T,P)$ from high-pressure-limit information\cite{Allen:2012jc}. 
The AutoTST tool presented here complements this approach by providing these high-pressure-limit reaction rates (except for barrierless reactions that will require a variational TST treatment), as well as direct calculations of barrier heights and vibrational frequencies of the transition states. 
Without these, the microcanonical rate coefficients $k(E)$ must be estimated via inverse Laplace transforms of canonical $k(T)$ rate expression estimates\cite{Allen:2012jc}.

\subsection*{Methods}
In brief, the transition state geometries were predicted using a modified version of our previously published algorithm\cite{Bhoorasingh:2015dza} that combines group-additive prediction of interatomic distances with distance-geometry methods, then optimized, and confirmed to correspond to the correct reaction using an intrinsic reaction coordinate (IRC) calculation. 
Symmetry numbers were determined via point group using the SYMMETRY software\cite{Patchkovskii:2003oj} and canonical TST calculations were performed using CanTherm\cite{CanTherm:2016}.  The entire process is automated in Python using modules and classes from RMG-Py\cite{Gao:2016dk}.
Full details of the algorithm and methods are described below.

\subsubsection*{Automated geometry searches}
Reactant and product structures were located using the automated algorithm developed in RMG and described by \citeauthor{Magoon:2012hg}\cite{Magoon:2012hg}.
Transition state structures were located using a group contribution method that predicts transition state reaction center distances using training data of known transition states. 
A brief explanation of the algorithm follows, with more information in ref.\ \citenum{Bhoorasingh:2015dza}.

The algorithm begins with generating upper and lower bounds of the distances between reactant atoms, known as a bounds matrix, which comes from a distance geometry approach in the open-source cheminformatics program RDKit \cite{RDKitOpensourcec:nApUqBGo}.
The bounds matrix is then edited for the transition state, whose reacting atom distances are estimated via the group contribution method. 
For hydrogen abstraction, these atom distances are those between the abstracted hydrogen, the atom bonded to the abstracted hydrogen, and the radical abstracting the hydrogen.

Molecular structure groups defining the reacting functional groups are organized into a hierarchical tree for each reaction family.
The tree was arranged based on chemical intuition of which molecular structure features are most important, i.e.~the number of radical electrons on a reacting atom is more important than its bonding configuration.
A training set for these reaction center distances was created from previously optimized transition state geometries.
The values for each group are calculated by finding a best fit to the training set, using linear least squares regression.
When estimating unknown distances for a transition state, the tree is traversed to the most specific group matching the molecular structure of the reacting groups.
The distance values for this group are added to base values stored in the top level of the tree, to provide all the reaction center distances needed for the reaction.

Once an estimate for the reaction center distances is made, the atoms are embedded in 3D space to create several conformers that satisfy the overall bounds matrix. 
These geometries are optimized, with the lowest energy conformer chosen as the transition state estimate, and can then be further optimized at a chosen level of theory. 
In this work, we use the M06-2X\cite{Zhao:2006dj,Zhao:2008kw} functional with a MG3S\cite{BenjaminJLynch:2003cy} basis set in  Gaussian 09\cite{Gaussian:we}.
The transition state is validated using a intrinsic reaction coordinate (IRC) calculation at the same level of theory;
the atomic coordinates from either end of the IRC pathway are converted into molecular graphs using a connect-the-dots algorithm\cite{Bhoorasingh:2015dza} and compared to the original reactants and products to confirm that the saddle point corresponds to the desired reaction.

%\paragraph*{Modifications to the group contribution transition state search}

\begin{figure}[H]
\centering	
\includegraphics[width=3.33in]{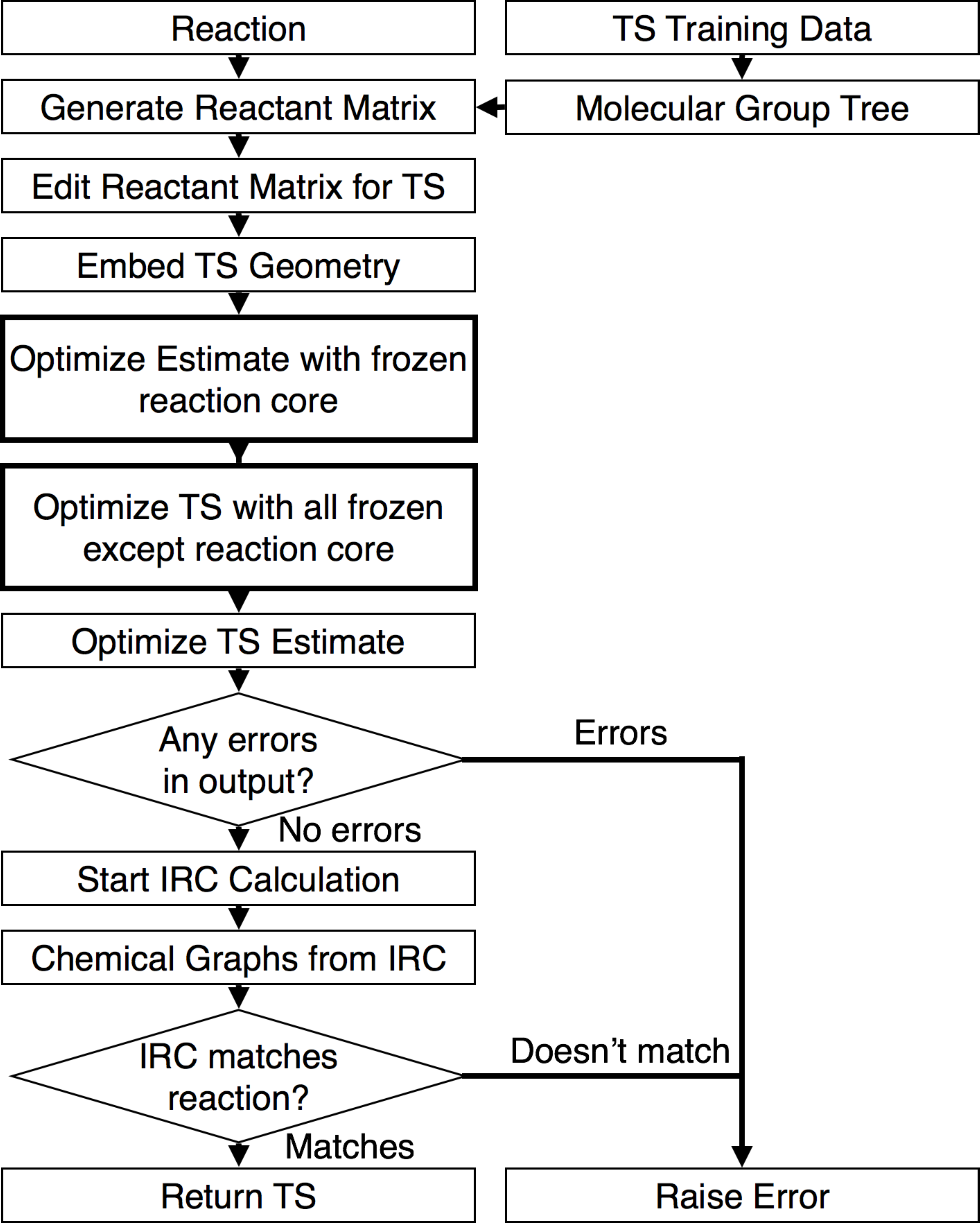}
\caption{\label{fig:mod_ts_algo} Automated transition state search algorithm. The steps with bold borders are deviations from the original algorithm described in ref.\  \citenum{Bhoorasingh:2015dza}.}
\end{figure}

The group contribution method for predicting transition state geometries as described in ref.\ \citenum{Bhoorasingh:2015dza} has been modified from its original formulation to improve its performance; the updated algorithm is shown in Figure \ref{fig:mod_ts_algo}.
The distance geometry algorithm in RDKit\cite{RDKitOpensourcec:nApUqBGo} requires upper and lower limits for the distances between every atom.
The difference between the upper and lower limits for the reaction center distances were previously set to 0.05\AA, but this was decreased to 0.025\AA~due to increased confidence in the reaction center predictions. 
Conformers were constructed in 3D to satisfy the distance limits for every atom pair.

The optimization protocol was also modified, with the transition state geometry prediction algorithm no longer using a universal force field optimization, instead adopting a protocol similar to that used in the AARON code \cite{Rooks:2014kz}.
The geometry estimate undergoes a constrained optimization to an energy minimum with the reaction center distances frozen,
followed by a transition state (saddle point) search with all distances frozen except the reaction center.
The resulting geometry is then used for a Berny transition state optimization\cite{JCC:JCC540030212}.

The transition state training data used in this study were optimized and validated at M06-2X with a MG3S basis set, so that predictions were made for the same electronic structure method used in this study.
M06-2X/MG3S provides sufficiently accurate kinetic parameters at a reasonable computational cost, and is widely available in computational chemistry packages\cite{Frisch:2009wv,Valiev:2010bb,Neese:2012ki}.
Our algorithm, previously demonstrated for hydrogen abstraction reactions, was modified and extended to be applied to intramolecular hydrogen migration and radical addition to multiple bond (\textbeta-scission in reverse) reaction families.

\subsubsection*{Kinetic calculations}
\begin{figure}[H]
\centering	
\includegraphics[height=4in]{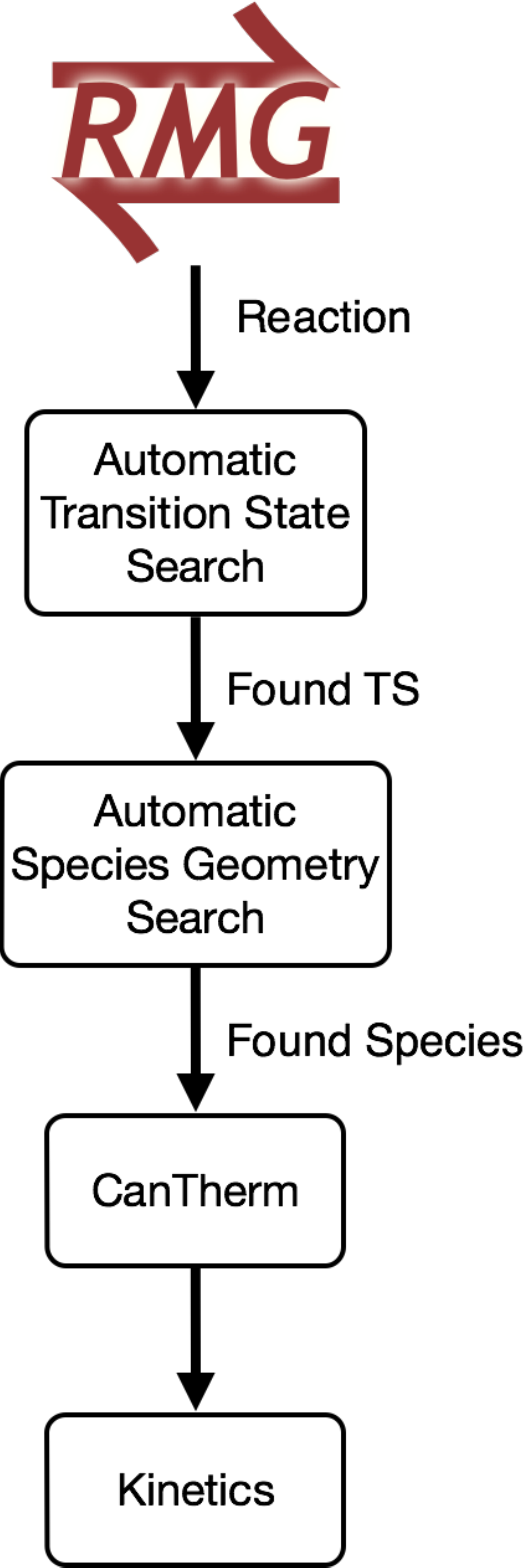}
\caption{\label{fig:kin_algo} The automated kinetic calculations involve an automated transition state search (Figure \ref{fig:mod_ts_algo}), automated search for reactant and product geometries\cite{Magoon:2012hg}, and automatically calculating kinetics using CanTherm\cite{CanTherm:2016}.}
\end{figure}

The CanTherm software package was used to determine kinetic parameters using classical transition state theory\cite{CanTherm:2016}.
Symmetry numbers for the rate calculations were determined via point group using the SYMMETRY software\cite{Patchkovskii:2003oj}.
SYMMETRY takes as input the optimized 3-dimensional geometry and a tolerance to allow for small deviations, and calculates the point group.
The point group is used to determine the symmetry number\cite{Irikura:1998ur}, and a chirality contribution is added for point groups that lack a superimposable mirror image.
Product geometries and energies were also found for these calculations so the Eckart model could be applied to determine tunneling corrections\cite{Eckart:1930kz}.
For the automated calculations the Rigid Rotor Harmonic Oscillator (RRHO) approximation is used for all reactants, products, and transition states.
Figure \ref{fig:kin_algo} provides an overview of the automated kinetic calculation method.

\subsubsection*{Comparison of Automated TST calculations and Rate Rules}
The AutoTST calculations were compared to rates
 of hydrogen abstraction, intramolecular hydrogen migration, and radical addition to multiple bond (and \textbeta-scission) reactions from the LLNL butanol combustion model\cite{Sarathy:2012fj},
 and to estimates of these rates generated by the automated rate rule implementation in RMG.
 Many of the rates in the hand-curated LLNL model were estimated by applying rate rules, while others were calculated or determined by analogy.
In RMG kinetics estimation, rates are determined by traversing hierarchical trees of rules based on molecular structure\cite{Gao:2016dk,RMGpy:documentation}.
If this results in an exact match for the specific molecular structure of the reactant(s), with that combination of nodes in the trees containing data, then this rule is used.
However, if the node exactly matching the reacting molecular structures does not contain data, the estimator will ``fall up'' to more general nodes, i.e. less precisely defined descriptions of the reactant(s). 

These more general nodes will be at some distance from the specific reactants' nodes as measured by the number of levels in the tree. Because there are several trees (e.g. one for each reactant), the Euclidean distance is calculated.
This ``Nodal distance'' can therefore be interpreted as the inappropriateness of the rate estimate used for each reaction, with an exact match corresponding to a Nodal distance of 0.

Kinetics were compared at 1000K, since the rate rules were determined for a combustion model in that temperature range.
For pressure-dependent rates in the LLNL model, the high pressure limits were used.

\subsubsection*{Comparison to benchmark calculations}
As will be seen in the Results section,
in some cases there are large differences between rate rule predictions and AutoTST calculations.
For two such cases from each reaction family, more thorough theoretical calculations were performed. 
The reactions were selected if there was at least a hundredfold discrepancy between the rate calculated by AutoTST and both the LLNL and RMG rates at 1000~K.
These benchmark calculations were compared to both the rate rule predictions and the AutoTST rates.

The geometries for the benchmark calculations were determined using the same DFT functional and basis set as the automatically calculated rates, but the benchmark calculations used an {\sl ultrafine} grid.
For the benchmark calculations, the 1-D hindered rotor approximation\cite{Pfaendtner:2007kv} was applied to torsional modes, instead of the RRHO approximation. 
AutoTST did not always find the lowest energy conformer, so when the hindered rotor scans revealed a lower-energy conformer this was re-optimized and adopted for the benchmark calculations.
Barrier heights were also recalculated using single point coupled-cluster calculations (see the `Computational chemistry' section for details).
Finally, symmetry numbers were manually checked, and corrected if the AutoTST approach was in error.

These improvements allowed comparison between AutoTST and the benchmark calculations to identify the sources of error in the AutoTST calculations.
To distinguish between the sources of error, each of the four improvements on the AutoTST method was systematically removed.
This analysis requires five manual quantum chemistry calculations in total for each reaction chosen, plus the automated calculation, so was only performed on six reactions not all 781 of the reactions successfully calculated with AutoTST.

\subsubsection*{Computational chemistry}
Geometry optimization and path analysis calculations used the M06-2X DFT functional\cite{Zhao:2006dj,Zhao:2008kw} with the MG3S basis set\cite{BenjaminJLynch:2003cy} (equivalent to 6-311+G(2df,2p) for systems containing C, H, and O)\cite{Clark:1983gu,Frisch:1984ep} in the Gaussian 09 quantum chemistry package\cite{Gaussian:we}.
For benchmark calculations, electronic energies were computed using the CCSD(T)-F12/RI method with the cc-VTZ-F12\cite{Peterson:2008eo} and cc-VTZ-F12-CABS\cite{Yousaf:2008jw} basis sets in ORCA\cite{Neese:2012ki}.

\section*{Results}
%\section{Comparison with rate-rule estimates}
Kinetics calculated with the integrated AutoTST algorithm were compared to two sets of estimates:
the first were from a butanol combustion model\cite{Sarathy:2012fj} from the Lawrence Livermore National Laboratory (LLNL) which used a combination of literature data, estimates, analogies, calculations, and rate rules; and the second set were rate rule predictions generated by RMG.

The butanol combustion model contained 855 hydrogen abstraction, 78 intramolecular hydrogen migration, and 184 radical addition to multiple bond reactions (including some reactions in the reverse \textbeta-scission direction).
For each reaction family, AutoTST calculated kinetics for approximately 70\% of the reactions (Table \ref{table:ModelRxns}).
The percentage successfully calculated was consistent across all reaction families, so the AutoTST success rate is so far independent of the reaction type.
Failures are usually due to the automated transition state search finding a saddle point that corresponds to a conformational change or a transition other than the intended reaction, and have been previously discussed\cite{Bhoorasingh:2015dza}.

\begin{figure}[H]
	\begin{flushleft}
	\begin{subfigure}[b]{\textwidth}
    \includegraphics[width=\textwidth]{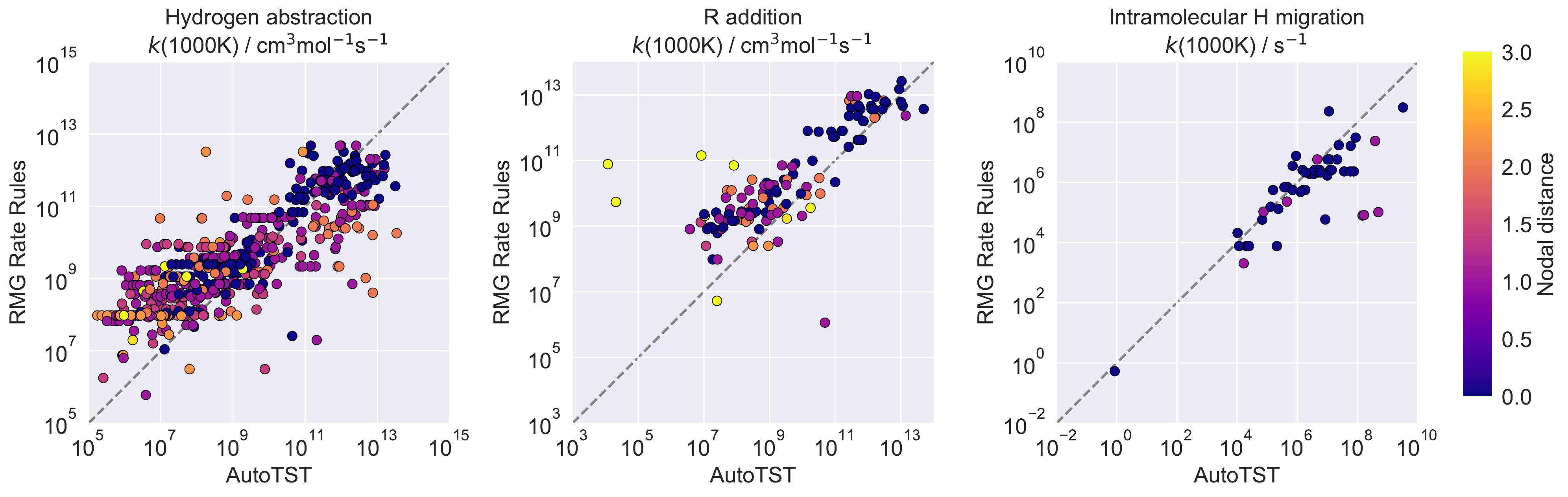}
    \caption{\label{fig:comp_RMG}}
	\end{subfigure}
    ~
    \begin{subfigure}[b]{0.94\textwidth}
    \includegraphics[width=\textwidth]{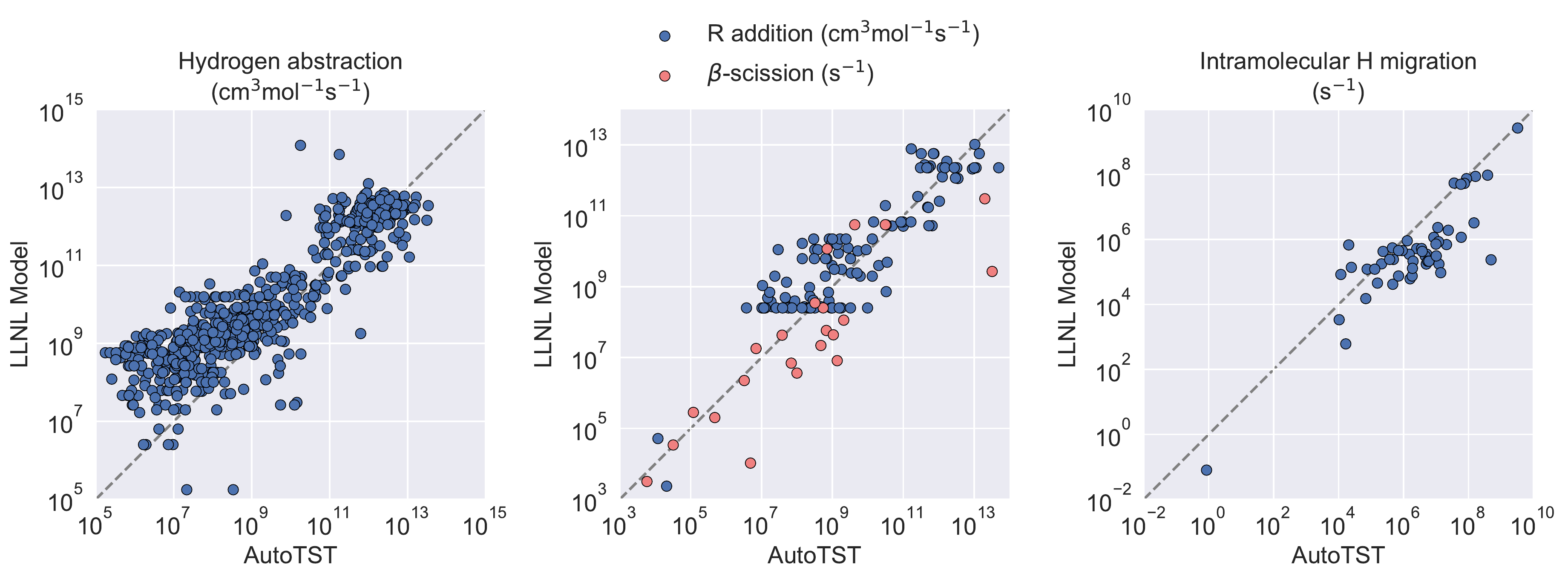}
    \caption{\label{fig:comp_Sarathy}}
	\end{subfigure}
	\caption{Rate rule estimates from RMG (a) and LLNL model rates (b)  plotted against automated algorithm TST calculations evaluated at 1000~K. From ref. \citenum{Bhoorasingh2017data}.}
    \label{fig:KinComp_all_compare}
    \end{flushleft}
\end{figure}

Overall, the AutoTST kinetics corresponded with with rate rule predictions from both sources (Figure \ref{fig:KinComp_all_compare}), with most rate rules being within an order of magnitude (factor of ten) of each other.
Figure \ref{fig:comp_RMG} displays the RMG rate rule predictions vs. AutoTST, with the color of each point corresponding to its nodal distance, a measure of inappopriateness of the rate rule used by RMG (see Methods section). 
The data and scripts used to produce Figure \ref{fig:comp_RMG} are freely available\cite{Bhoorasingh2017data}.

For both hydrogen abstraction and radical addition to multiple bond reactions, the AutoTST rates that matched best with the RMG rates, generally, were for rates with low nodal distances.
Where significant ``banding'' is seen on the plots, RMG is using the same reaction rate for several reactions; these points are mainly in orange indicating a higher nodal distance. 
In the intramolecular hydrogen migration reactions, the nodal distances never exceed 1 and less scatter is seen than for the other reaction families.
Using the nodal distance to estimate errors in the rate rule estimates could guide when it is appropriate to use AutoTST calculations in the context of automated mechanism generation using RMG. 

AutoTST agreement with LLNL rates is similar to its agreement with RMG rate rules, as indicated by Figure \ref{fig:comp_Sarathy}. 
Banding in the AutoTST vs. LLNL plots for hydrogen abstraction and radical addition to multiple bond, where the same rate is being used for 10 or more reactions in some cases, illustrates that this published model also applies rate rules and analogies with varying degrees of applicability.
Both sets of comparisons indicate that AutoTST agrees with rate rules when the rules are applied appropriately, but provides a better method of calculation in the absence of good rate estimates.

The Gaussian 09 log files and CanTherm input and output files for all 781 reactions and 282 corresponding species are available online in a FigShare repository\cite{Bhoorasingh2016data}.

\begin{table}[H]
\small
\caption{\label{table:ModelRxns} Number of reactions for each family contained in the combustion model, and success of the AutoTST algorithm.}
\centering
\begin{tabular*}{\linewidth}{@{\extracolsep{\fill}}lccc}
\hline\rule{0pt}{2.6ex}%
Reaction Family & Number of  & Kinetics successfully & Percentage \\
  & Reactions  & calculated & calculated\\
\hline\rule{0pt}{2.6ex}%
Hydrogen abstraction & 855 & 598 & 70\% \\
Intramolecular hydrogen migration & 78 & 52 & 67\% \\
Radical addition to multiple bond & 184 & 131 & 71\% \\
\hline\rule{0pt}{2.6ex}%
Total & 1117 & 781 & 70\%\\
\hline
\end{tabular*}
\end{table}

%\subsection*{Comparison of automated TST calculations and rate rules}

%\begin{figure}[H]
%\centering	
%\includegraphics[width=4in]{habs_parityplot}
%\caption{\label{fig:KinComp_HAbs} Rate rule estimates ($y$-axis)  plotted against automated algorithm TST calculations ($x$-axis) at 1000~K.}
%\end{figure}

%\vspace{1 em}
\section{Benchmark calculations}
\begin{figure}
	\centering
	\begin{subfigure}[b]{3in}%0.45\textwidth}
    \includegraphics[width=\textwidth]{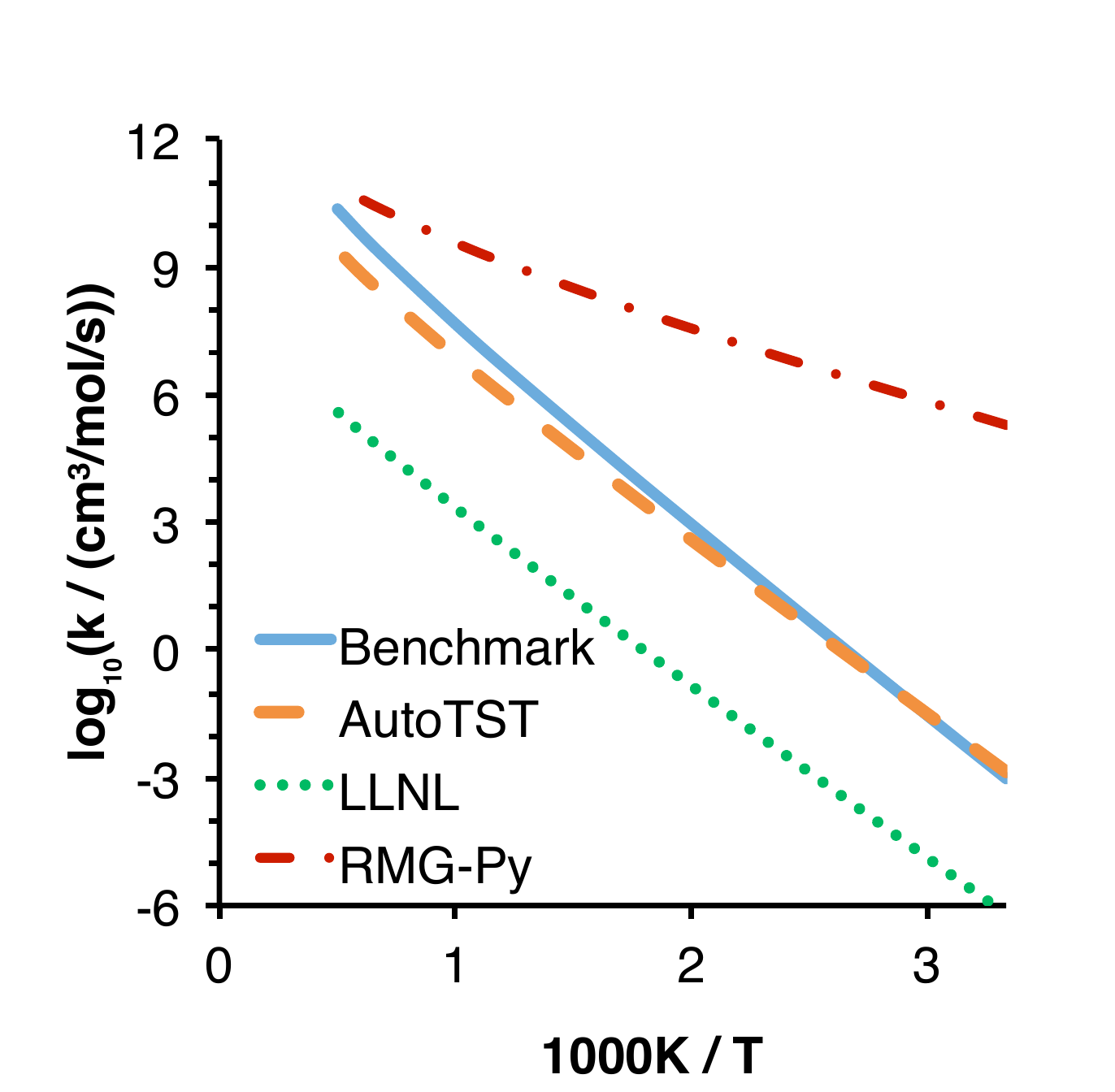}
    \caption{\label{fig:RAddMult_comp1}}
	\end{subfigure}
    ~
    \begin{subfigure}[b]{3in}%0.45\textwidth}
    \includegraphics[width=\textwidth]{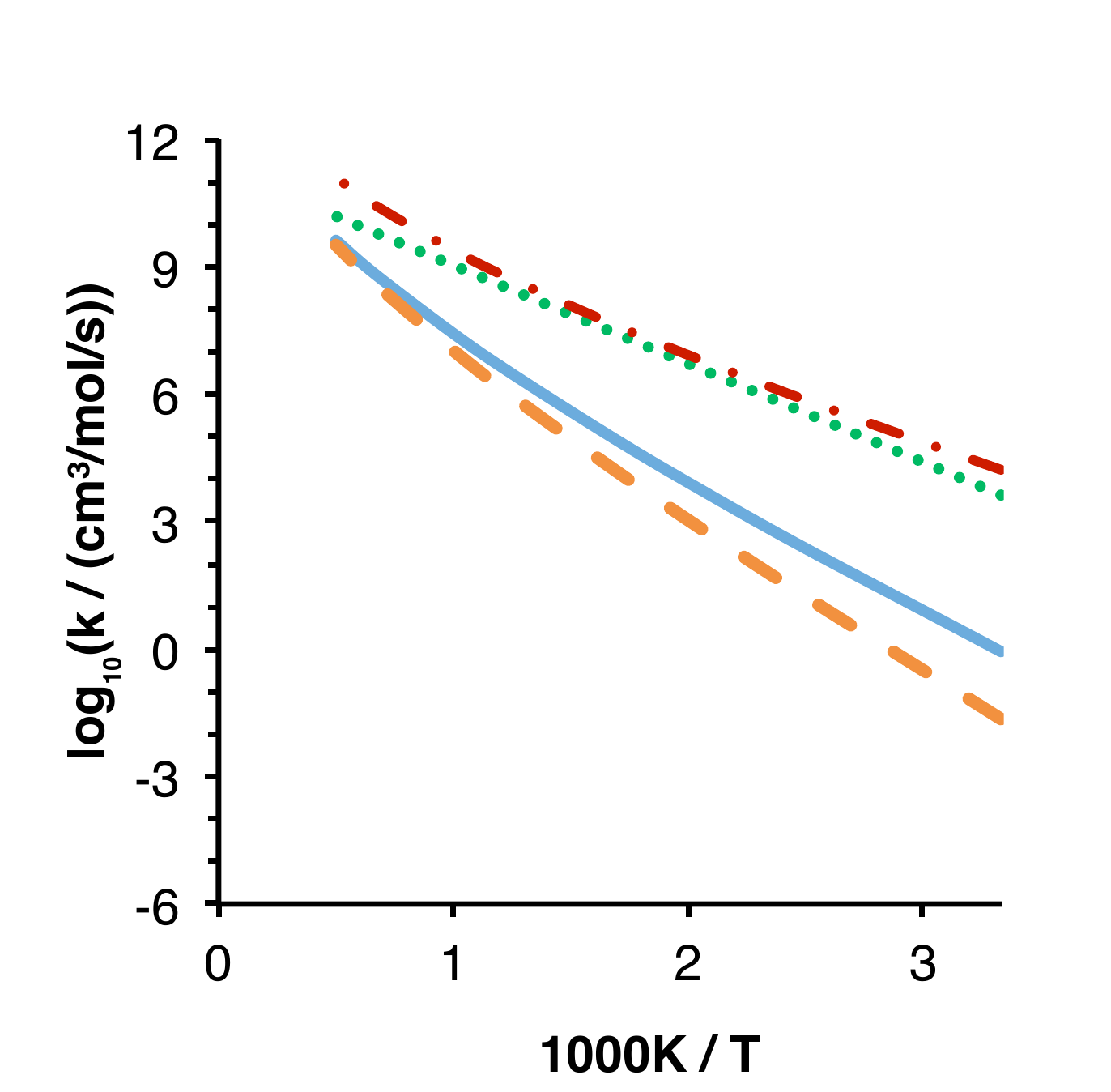}
    \caption{\label{fig:RAddMult_comp2}}
	\end{subfigure}
	\caption{Comparison of kinetic estimates for radical addition to multiple bond reactions R5 (a) and R6 (b).}
    \label{fig:RAddMult_Comp}
\end{figure}

Despite the overall agreement, a number of reactions had significant discrepancy between AutoTST rates and the rate rule predictions.
Six of the reactions with significant discrepancies, two from each reaction family (Table \ref{table:BenchRxns}), were selected for benchmark calculations as described above, to determine the accuracy of the three prediction methods and the sources of error.
For reaction 5, the rate from the LLNL model was provided in the reverse (\textbeta-scission) direction, so the rate shown (Figure  \ref{fig:RAddMult_comp1}) was calculated using the provided rate and thermodynamics from the model.
% Details of the benchmark calculations are included in the supporting information.
%The rate rule predictions were closer than  AutoTST to  the benchmark calculations for reaction 1 (Figure \ref{fig:habs_comp1}), but the AutoTST calculations showed better agreement than the rate rules for reactions 2 through 6 (Figure \ref{fig:habs_comp2} and supplementary materials).
%, \ref{fig:iHA_Comp}, and \ref{fig:RAddMult_Comp}).

%\subsection*{Comparing predictions to benchmark calculations}

\begin{table}[htbp]
\small
\caption{\label{table:BenchRxns} Reactions compared to benchmark calculations.}
\centering
\begin{tabular*}{\linewidth}{@{\extracolsep{\fill}}c l c}
\hline\rule{0pt}{2.6ex}%
Label & Family & Reaction \\
\hline\rule{0pt}{2.6ex}%
R1 & H abstraction & \ce{C2H5OO^. + C2H6 <=> C2H5OOH + ^.CH2CH3} \\
R2 & H abstraction &  \ce{^.OOH + CH3C(=O)C2H5 <=> H2O2 + ^.CH2C(=O)C2H5} \\
R3 & Intramolecular H migration & \ce{O=CHCH2OO^. <=> O=C^.CH2OOH} \\
R4 & Intramolecular H migration & \ce{CH3C(CH3)(C=O)OO^. <=> CH3C(CH3)(^.C=O)OOH} \\
R5 & Radical addition & \ce{CO2 + ^.CH3 <=> CH3C(=O)O^.} \\
R6 & Radical addition & \ce{CH2C(CH3)CH=O + HO2^. <=> ^.CH2C(CH3)(CH=O)OOH} \\
\hline
\end{tabular*}
\end{table}

\begin{figure}[H]
	\centering
	\begin{subfigure}[b]{3in}%0.45\textwidth}
    \includegraphics[width=\textwidth]{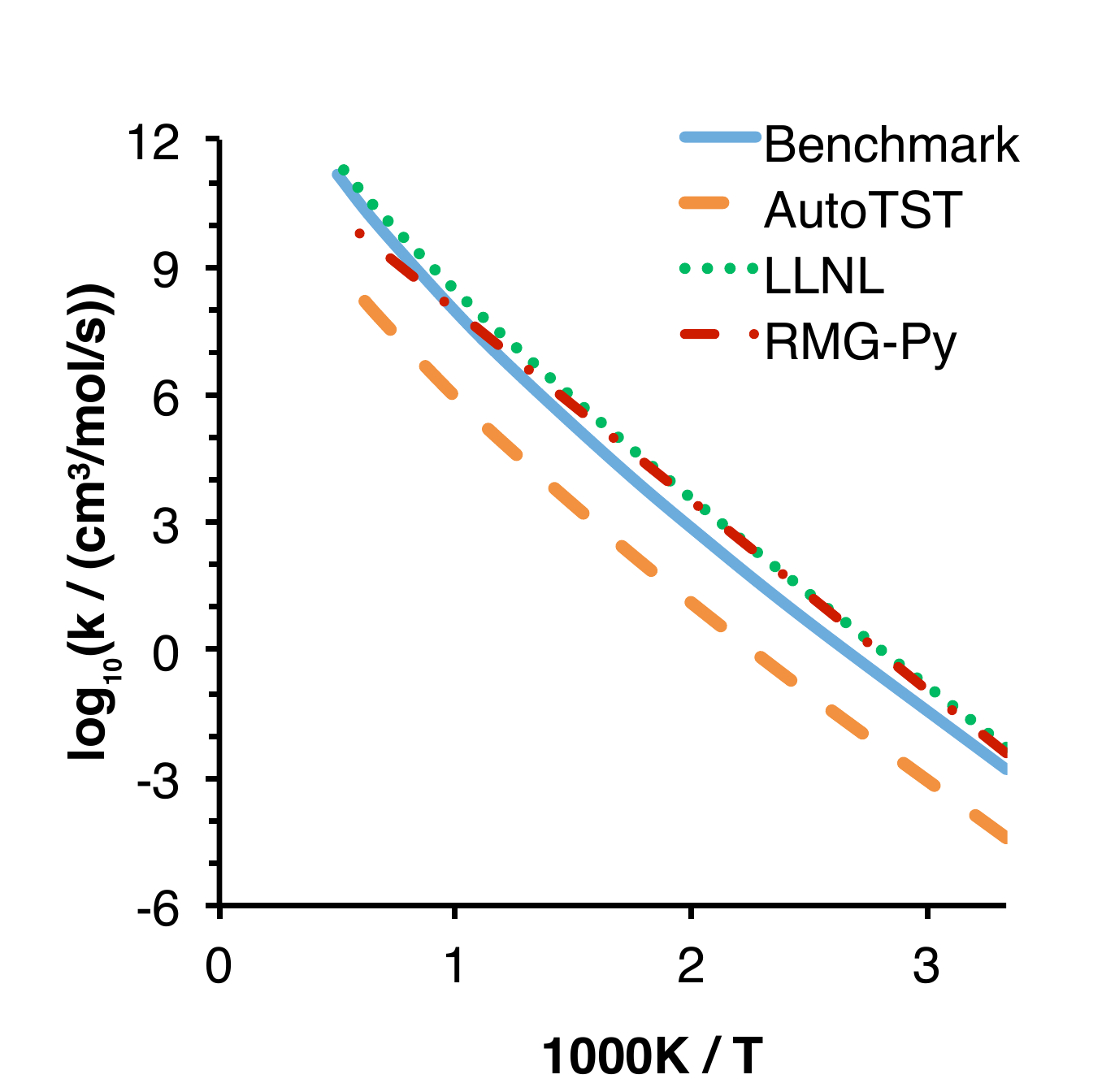}
    \caption{\label{fig:habs_comp1}}
	\end{subfigure}
    ~
    \begin{subfigure}[b]{3in}%0.45\textwidth}
    \includegraphics[width=\textwidth]{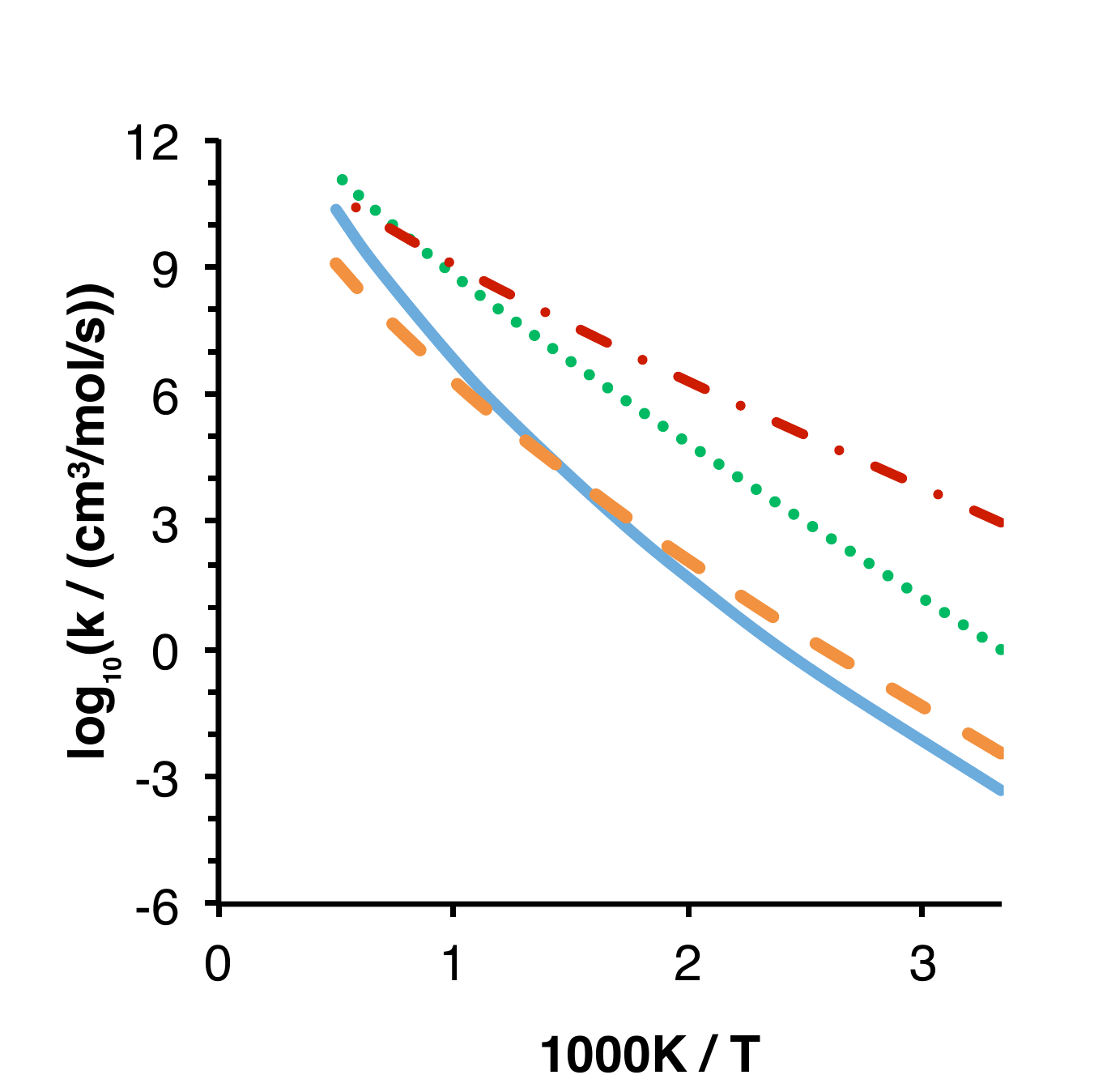}
    \caption{\label{fig:habs_comp2}}
	\end{subfigure}
	\caption{Comparison of kinetic estimates for hydrogen abstraction reactions R1 (a) and R2 (b). }
	% Well defined rate rules exist for reaction 1, resulting in the predictions outperforming the automated calculations. The automated calculations outperformed the rate rule predictions for reaction 2.
    \label{fig:HAbs_Comp}
\end{figure}

\begin{figure}
	\centering
	\begin{subfigure}[b]{3in}%0.45\textwidth}
    \includegraphics[width=\textwidth]{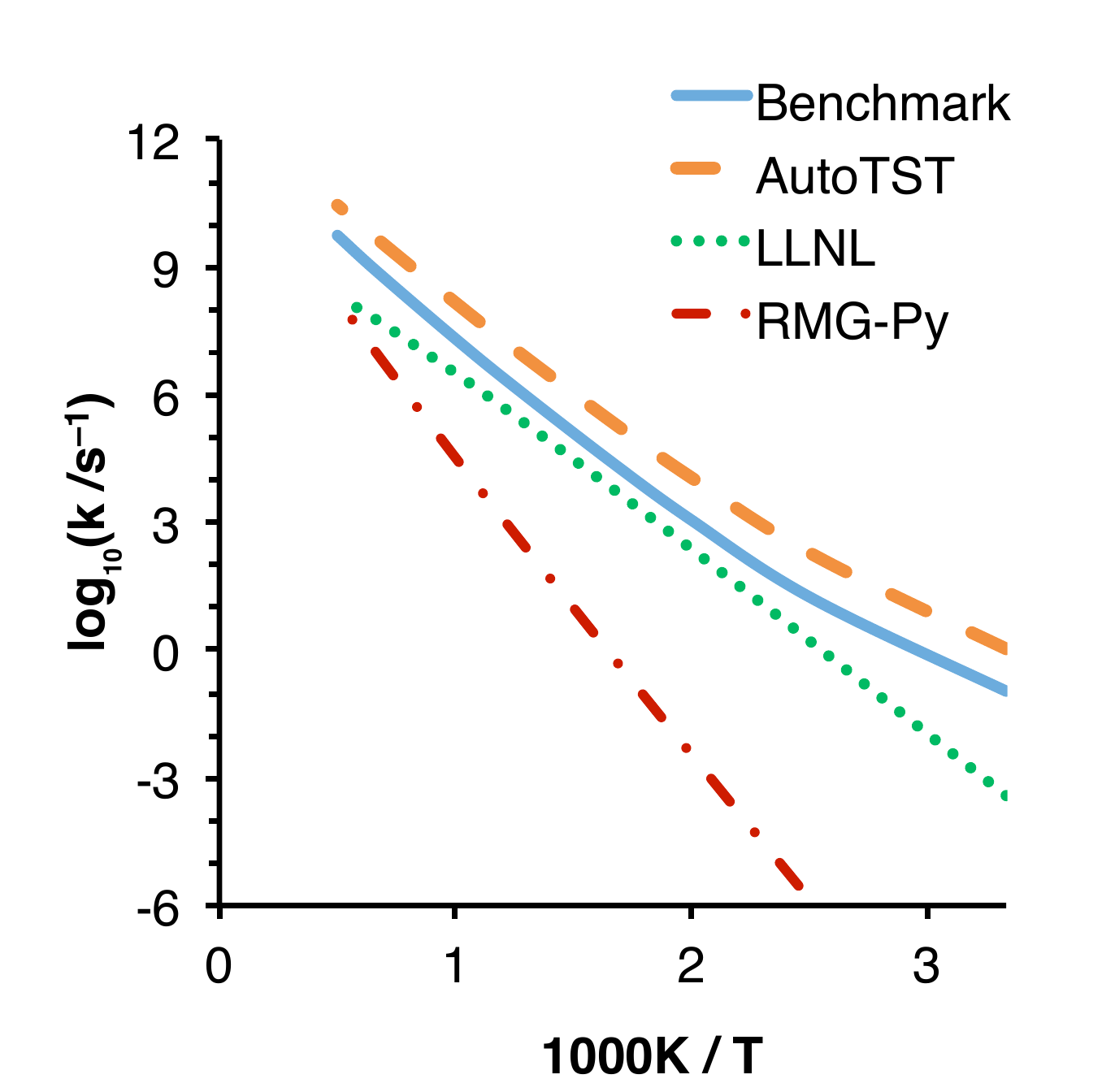}
    \caption{\label{fig:iha_comp1}}
	\end{subfigure}
    ~
    \begin{subfigure}[b]{3in}%0.45\textwidth}
    \includegraphics[width=\textwidth]{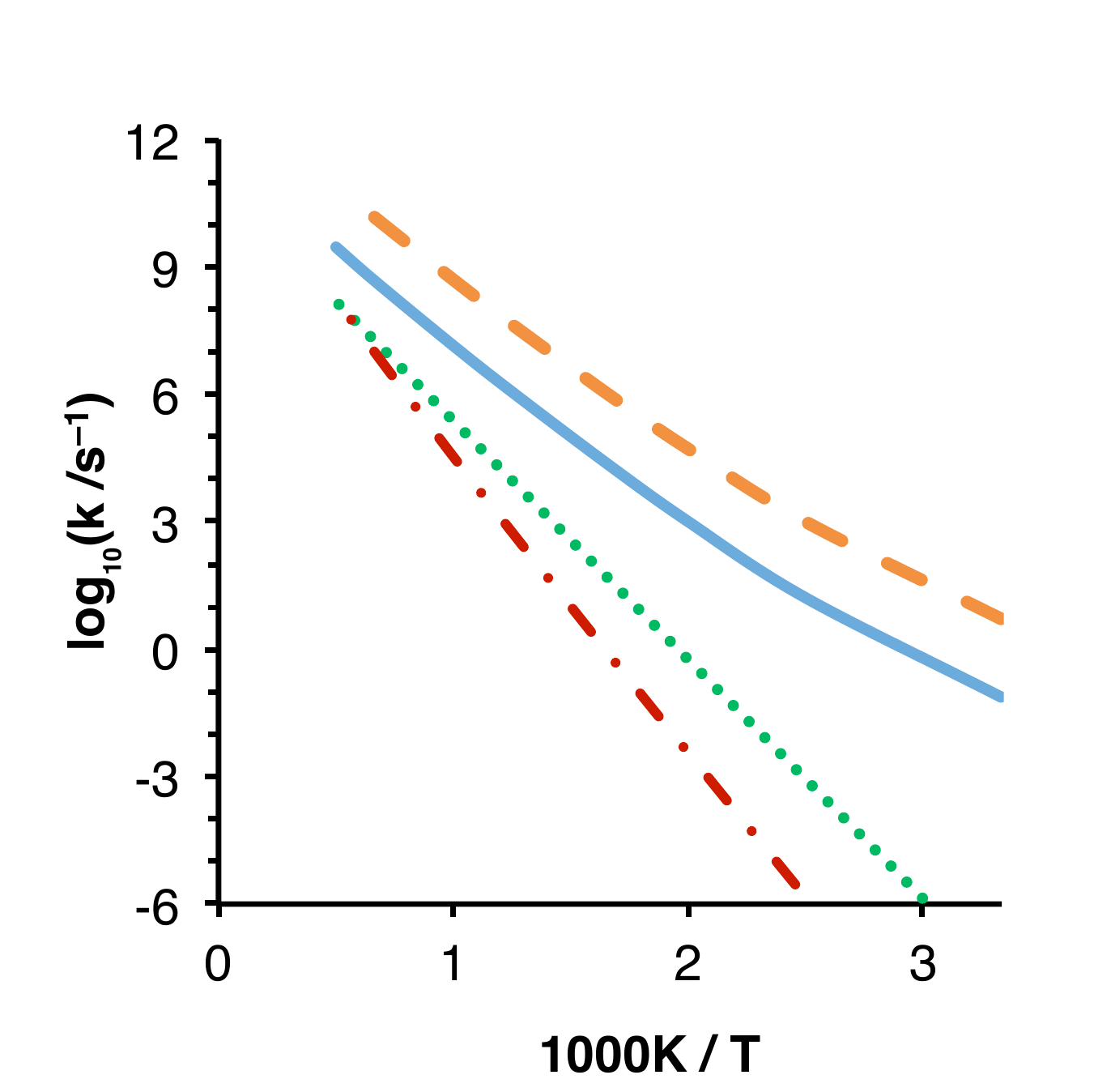}
    \caption{\label{fig:iha_comp2}}
	\end{subfigure}
	\caption{Comparison of kinetic estimates for intramolecular hydrogen migration reactions R3 (a) and R4 (b).}
    \label{fig:iHA_Comp}
\end{figure}

For one case (reaction 1, Figure \ref{fig:habs_comp1}), both rate rule estimates outperformed the AutoTST calculation, in that they were closer to the benchmark rate.
The rate rule used in the LLNL model was developed for \ce{RO2^. + C2H6 <=> ROOH + C2H5}\cite{Carstensen:2007ks}, and the RMG rate rule was developed for \ce{HO2^. + C2H6 <=> H2O2 + C2H5}\cite{Walker:1989fv}, both of which are quite similar to R1 (\ce{C2H5OO^. + C2H6 <=> C2H5OOH + C2H5}).
The accuracy of the rate rule predictions should be expected since the rules were developed for reactions resembling R1, and in such cases AutoTST should not be used since the rate rules could provide a good rate prediction at a far lower computational cost.

AutoTST outperformed the rate rule predictions for all the other reactions (Figures \ref{fig:habs_comp2}, \ref{fig:iHA_Comp}, and \ref{fig:RAddMult_Comp}).
In one such example (reaction 3, Figure \ref{fig:iha_comp1}), the RMG rate prediction was made from a generalized rate rule (high nodal distance), so the kinetic data used to make the prediction was very unlike the reaction, leading to the large discrepancy in the reaction rate.
The value used in the LLNL model was not an estimate but a theoretically calculated value for the specific reaction \cite{Lee:2003kt}, which agrees favorably with the benchmark calculation.

Overall, the comparison of the 6 reactions with large discrepancies between rate rule predictions and AutoTST show that the automated method performs well for all tested reaction families.
This is particularly true when considering the performance of the kinetics across a wide temperature range, where the kinetics calculated with AutoTST agree with the high accuracy calculations, but the rate rules become less accurate outside the temperature range for which they were developed.
%In other such cases, when specific reaction rates are available, the available data should be relied on, bypassing AutoTST.

%\subsection*{Sources of error}
Discrepancies between the AutoTST calculations and the benchmark calculations presented an opportunity to identify sources of error in the AutoTST method:
using the DFT energies instead of CCSD(T)-F12/RI, neglecting hindered rotors, using an automatically determined symmetry number that may be incorrect, and using randomly guessed conformers which may not be the lowest energy choices.
To isolate the effects of each error, each correction was individually removed from the benchmark calculations and replaced with the equivalent used for the automated calculations.

\begin{figure}
\centering	
\includegraphics[width=\textwidth]{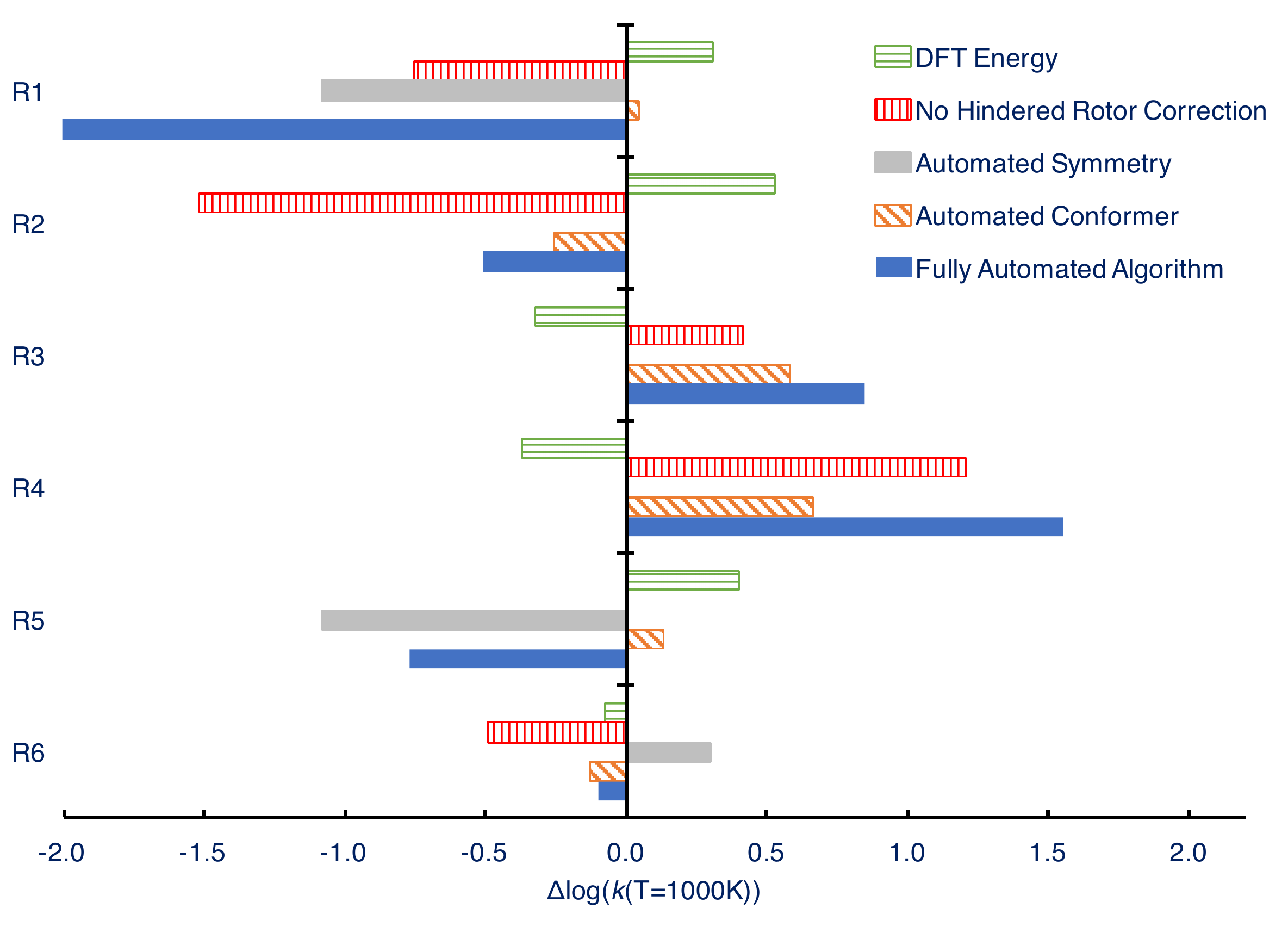}
\caption{\label{fig:k1000_error} Sources of error in each automated algorithm compared to its respective benchmark calculation. $\Delta \log(k(1000\text{ K})) = \log_{10}\left((k \text{ with a correction omitted})/(\text{benchmark } k\text{ with all corrections})\right)$. Omitting all the corrections simultaneously gives the ``Fully Automated Algorithm'' result.}
\end{figure}

Figure \ref{fig:k1000_error} shows the magnitude of the difference in the rate coefficient at 1000~K due to each source of error.
Table \ref{table:Activation_Energy} displays the difference in activation energy due to each source of error and the benchmark calculation, and Table \ref{table:A_Factor} shows the changes to the pre-exponential ``$A$ factor'' due to the same effects.
%Tables displaying the difference in both the activation energy and the pre-exponential ``$A$ factor'' due to the same effects are included in the supporting information.

\begin{table}[H]
\caption{\label{table:Activation_Energy} Difference in the activation energy (kJ/mol) compared to the benchmark calculations. Kinetics fitted to Arrhenius form between 600K and 2000K.}
\centering
\begin{tabular*}{\linewidth}{@{\extracolsep{\fill}}|c|c|cccc|c|}
\hline\rule{0pt}{2.6ex}%
Reaction & Benchmark  & Inaccurate & Hindered & Incorrect & Incorrect & Overall\\
  & E$_a$  & Energy & Rotors & Symmetry & Conformer & Discrepancy\\
\hline\rule{0pt}{2.6ex}%
R1 & 111.98 & --5.46 & --0.34 & 0.00 & --0.91 & --6.01\\
R2 & 119.38 & --9.92 & --20.70 & 0.00 & +4.57 & --24.53\\
R3 & 88.54 & +6.16 & +0.28 & 0.00 & --11.05 & --4.06\\
R4 & 85.19 & +7.03 & +1.08 & 0.00 & --12.52 & --3.35\\
R5 & 97.24 & --7.73 & 0.00 & 0.00 & --2.52 & --7.68\\
R6 & 84.89 & --6.74 & --8.35 & 0.00 & +0.01 & +1.91\\
\hline
\end{tabular*}
\end{table}

\begin{table}[H]
\caption{\label{table:A_Factor} Difference in the log$_{10}$ of the $A$ factor compared to the benchmark calculations. Kinetics fitted to Arrhenius form between 600K and 2000K. R3 and R4 are in [s\textsuperscript{-1}] and the others are in [cm$_3$/(mol s)].}
\centering
\begin{tabular*}{\linewidth}{@{\extracolsep{\fill}}|c|c|cccc|c|}
\hline\rule{0pt}{2.6ex}%
Reaction & Benchmark &Inaccurate & Hindered & Incorrect & Incorrect & Overall\\
 & log$_{10}$A  & Energy & Rotors & Symmetry & Conformer & Discrepancy\\
\hline\rule{0pt}{2.6ex}%
R1 & 13.96 & +0.020 & --0.772 & --1.079 & 0.000 & --2.337\\
R2 & 13.27 & +0.010 & --2.659 & 0.000 & -0.016 & --1.854\\
R3 & 12.00 & --0.001 & +0.416 & 0.000 & +0.004 & +0.620\\
R4 & 11.64 & --0.002 & +1.253 & 0.000 & +0.004 & +1.369\\
R5 & 12.83 & 0.000 & 0.000 & --1.079 & 0.000 & --1.169\\
R6 & 11.53 & 0.000 & --0.483 & 0.000 & 0.000 & +0.141\\
\hline
\end{tabular*}
\end{table}

The major source of error for AutoTST calculations was using the RRHO approximation for treatment of internal rotations, although not for reactions that contained few rotors (e.g. R5).
Symmetry was also a major source of error when the automated method determined symmetry numbers incorrectly.
This was not consistent for all tested reactions as the automated method correctly determined symmetry for some cases.
As expected, the activation energy is unaffected by correcting the symmetry number but is changed somewhat by using coupled-cluster calculations instead of DFT for barrier heights.
%(Table \ref{table:Activation_Energy}).
Correcting the DFT energy had little effect on the rate calculations in the combustion temperature range, but at lower temperatures the DFT energy led to rates that were approximately an order of magnitude different from the benchmark calculations.
AutoTST was not always successful in finding the lowest energy conformer for all structures, which contributed to errors of varying degrees.
The intramolecular hydrogen migration reactions were most affected by these, where a single wrong conformer would contribute significantly to an error in the barrier height.

\section*{Discussion}

While all sources of error need to be addressed, automating the treatment of hindered internal rotors and providing a more robust algorithm for determining symmetry should be targeted first.
Automating hindered rotor calculations will also help to identify the lowest energy conformer, thus correcting errors due to the incorrect conformer selection.
%The current major sources of error in the automated kinetics are improperly determining symmetry, and not accounting for internal rotor contributions. Based on these errors, the automated algorithm can be further improved to provide even better parameters.
%Improved automated methods for determining symmetry numbers would reduce the uncertainty in AutoTST calculations.
%Internal rotor contributions could be included in the calculations by automating hindered rotor calculations.
%The hindered rotor calculations would also help identify an existing lower energy conformer, correcting cases where the lowest energy conformer was not automatically found.
The additional computational cost of more accurate electronic structure calculations to improve barrier heights would have to be balanced against available computational resources, as using the DFT energy was not a major source of discrepancy.

\section*{Conclusion}

%AutoTST has calculated kinetics for approximately 70\% of all tested reactions.
In summary, the AutoTST method
%The method
 is extensible, and has now been applied to hydrogen abstraction, intramolecular hydrogen migration or transfer, and radical addition to multiple bond reactions.
The successful extension of AutoTST motivates further work to include other reaction families with a reaction barrier.
%To complete a kinetic model the method will need extending to additional reaction families.
Since the method is shown to work for all three main classes (isomerization, decomposition, bimolecular), extending to additional reaction families is mostly a case of training new group values to predict the interatomic distances, as described in ref.\ \citenum{Bhoorasingh:2015dza}. 

Despite current sources of error, good kinetic estimates can be calculated using the automated algorithm, and the protocol should be used for mechanism generation alongside other rate estimation methods.
The algorithm can also be used as a stand-alone tool to obtain specific reaction rates or transition state geometries.
Further, the tool can be used semi-manually to set up a transition state geometry calculation with good interatomic distances, and the geometry can be optimized further and used for kinetics and other calculations at the user's desired level of theory.

AutoTST outperforms rule-based kinetic estimation methods when specific rate rules are unavailable for a reaction.
The current estimation methods should not be abandoned, however, as they can still provide good kinetic predictions when used appropriately (e.g.\ R1), and in a computationally efficient manner.
A new module in RMG allows the user to obtain an estimate for the uncertainty of a reaction rate.
When rates from estimation methods are unavailable, or the rates prove too uncertain to be used confidently in model generation, AutoTST now provides an alternate method to determine kinetics with little human input.
%Despite the current sources of error, the method can provide improved kinetic parameters for many reactions in microkinetic models, reducing the uncertainty of these models. 
The fully automated process represents an important step towards high-throughput calculation of many thousands of accurate reaction rates, which can now be feasibly used in the construction of detailed kinetic models. %which until now required too much human input to be feasible.
%AutoTST should be used to target reactions where the kinetics are estimated with more generalized rate rules.
Another potential use of this tool is for generating large quantities of training data as input to other prediction algorithms.
\begin{acknowledgement}
The authors thank C. Franklin Goldsmith (Brown University), Jorge Aguilera Iparraguirre (Harvard University, now at Kyulux), and the developers of RMG at both Northeastern University and the Massachusetts Institute of Technology for helpful discussions. 
Acknowledgment is made to Northeastern University Department of Chemical Engineering and the Donors of the American Chemical Society Petroleum Research Fund for partial support of this research.

\end{acknowledgement}

%\begin{suppinfo}
\section*{Supporting Information Available}
% A listing of the contents of each file supplied as Supporting Information
% should be included. For instructions on what should be included in the
% Supporting Information as well as how to prepare this material for
% publications, refer to the journal's Instructions for Authors.

The following files are available free of charge.
\begin{itemize}
  \item The Gaussian 09 log files and CanTherm input and output files for all 781 reactions and 282 corresponding species are available in the figshare repository\cite{Bhoorasingh2016data} at 
 \\{https://doi.org/10.6084/m9.figshare.4234160}
  \item The results in CSV format and Python scripts required to generate Figure \ref{fig:KinComp_all_compare} are available in 
 the figshare repository \cite{Bhoorasingh2017data} at \\{https://doi.org/10.6084/m9.figshare.5244640}
\end{itemize}

%\end{suppinfo}

\bibliography{kinComp}
%\bibliography{achemso-demo} % remove this once no longer needed

\end{document}